\documentclass[review,12pt]{elsarticle}
\usepackage{color}
\usepackage{lineno,hyperref}
\modulolinenumbers[5]
\usepackage{amsmath,amsfonts,graphicx,epsfig}
\journal{Journal of \LaTeX\ Templates}
\newcommand{\ee}{\end{eqnarray}}
\newcommand{\be}{\begin{eqnarray}}

\begin{document}

\begin{frontmatter}

\title{\bf On primordial black holes from an inflection point}

\author{Cristiano Germani~\footnote{\tt email: germani@icc.ub.edu}}
\address{Institut de Ciencies del Cosmos (ICCUB), Universitat de Barcelona, Mart Franqus 1,
E08028 Barcelona, Spain}
\author{Tomislav Prokopec~\footnote{\tt email: t.prokopec@uu.nl}}
\address{Institute for Theoretical Physics, Spinoza Institute and EMME$\Phi$, Utrecht University,\\
Postbus 80.195, 3508 TD Utrecht, The Netherlands}

\begin{abstract}

Recently, it has been claimed that inflationary models with an inflection point in the scalar potential can produce a large resonance in the power spectrum of curvature perturbation. In this paper however we show that the previous analyses are incorrect. The reason is twofold: firstly, the inflaton is over-shot from a stage of standard inflation and so deviates from the slow-roll attractor before reaching the inflection. Secondly, on the (or close to) the inflection point, the ultra-slow-roll trajectory supersede the slow-roll one and thus, the slow-roll approximations used in the literature cannot be used.
We then reconsider the model and provide a recipe for how to produce nevertheless a large peak in the matter power spectrum {\it via}
fine-tuning of parameters.
\end{abstract}


\end{frontmatter}



\section{Introduction}

Undoubtedly, the most fascinating resolution of the missing Dark Matter (DM) problem would be that DM
is entirely formed by black holes generated during inflation, {\it i.e.} by primordial black holes (PBHs).
While until very recently it has been thought that this possibility is ruled out by Cosmic Microwave Background (CMB) observations, because of a significant numerical mistake in an earlier work~\cite{ricotti}, discovered in~\cite{kamionkowski}, this possibility has now come back in full glory. Although there are many possible astrophysical constraints on PBHs, those are questionable \cite{carrlast}, \cite{florian} and thus, until further solid astrophysical arguments are given,
one should seriously consider the possibility that PBHs constitute the whole (or most) of DM.

CMB observations are in full compatibility with simplicity: a stage of single field inflation is consistent with all observations.~\cite{planck}.
Therefore, although PBHs could in principle be formed in a multi-fields scenario~\cite{clesse}, it is natural to wonder
whether single field scenarios might also generate PBHs. Along this line of thought,
in an important paper 
Gacia-Bellido and Ruiz Morales~\cite{bellido} have recently investigated the idea which posited that a change of
curvature of the inflationary potential from a steeper, where CMB curvature perturbations are generated, to a shallower one, generate a large amplification of the power spectrum of curvature perturbations. This amplification can then trigger gravitational collapse and therefore PBH formation. The fascinating fact is that such a feature of the potential,
and indeed a potential that closely resembles the one of~\cite{bellido}, can naturally appear in Higgs inflation~\cite{yuta,critical},\cite{bellidohiggs}.

If inflation is in a slow-roll attractor, when the inflaton approaches an inflection point
(at which the first derivative of the potential vanishes), na\"ively one would expect that the amplitude of primordial
fluctuations would exhibit an infinite enhancement,
since the power spectrum is inversely proportional to the derivative of the potential.
This, of course, cannot be true. In reality, close to an inflection point (and also close to a near inflection point
used in~\cite{bellido}) the inflaton must either exit from inflation or enter an ultra-slow-roll
regime~\cite{martin,Tsamis:2003px,Kinney:2005vj,ultra,Mooij:2015yka,Romano:2016gop}. In such a  regime
the potential is flat and the friction due to the expansion is compensated for by the inflaton's deceleration,
slowing it down rapidly such that the universe approaches a de Sitter stage.

Although this fact invalidates immediately the analysis of~\cite{bellido}, at first sight the situation is not bad at all
as during an  ultra-slow-roll stage the spectrum grows exponentially.
Unfortunately though, unless parameters of the model studied in \cite{bellido} are appropriately tuned,
the exponential growth of the ultra-slow-roll phase is too short and yields no significant enhancement.
The reason is that, before entering into the ultra-slow-roll regime,
because of the change of curvature the inflaton shoots-over to a larger velocity that lies outside the standard inflationary
slow-roll attractor. During that transition the power spectrum decreases typically by several orders of magnitude
 so that the subsequent ultra-slow-roll amplification does not last long enough to even recover the lost magnitude.

In order to avoid the fatal overshooting and obtain a sizeable amplification of the power spectrum, one has to reduce the overshooting and have a long-lasting ultra-slow-roll phase. In what follows we give an example
in which both are realised and the amplification is effective.


\section{The overshoot}
\label{The overshoot}

Consider a flat Friedmann-Lema\^itre-Robertson-Walker (FLRW) spacetime with the metric,~\footnote{In this
work we use natural units, in which $c=1=\hbar$.}
\be
ds^2=-dt^2+a(t)^2 d\vec{x}\cdot d\vec{x}\, .
\label{metric}
\ee
A quasi-de Sitter stage needed for cosmic inflation is attained within a FLRW space~(\ref{metric})
whenever the principal slow-roll parameter,
\be
 \epsilon\equiv-\frac{\dot H}{H^2}\ll 1
\,,
\label{epsilon:g}
\ee
where $H\equiv \dot a/a$ is the Hubble parameter and
$\dot a = da/dt$.
The slow-roll attractor of standard inflation is reached
whenever both (geometric) slow-roll parameters, $ \epsilon$ and
\be
\epsilon_2 = \frac{\dot\epsilon}{\epsilon H}
\label{eta:g}
\ee
are small, {\it i.e.} when $\epsilon,|\epsilon_2|\ll 1$. On the other hand, on a slow-roll attractor, the slow-roll parameters satisfy,
\be
\epsilon \simeq  \epsilon_{\rm sr}\equiv\frac{M_p^2}{2}\frac{V'^2}{V^2}\, ,
\qquad
\epsilon_2  \simeq  4\epsilon_{\rm sr} - 2\eta_{\rm sr}
\,,\qquad \eta_{\rm sr}\equiv M_p^2\frac{V''}{V}\, ,
\label{slow-roll parameters: definition}
\ee
where $V(\phi)$ denotes the scalar field ($\phi$) potential, $V '= dV(\phi)/d\phi$,
$M_p=1/\sqrt{8\pi G_N}\simeq 2.2\times 10^{18}~{\rm GeV}$
is the reduced Planck mass and $G_N$ is the Newton constant.

The ultra-slow-roll phase is instead defined by a large $\epsilon_2$ but still a small $\epsilon$.
 Generically, in ultra-slow-roll,
\be
\epsilon \neq \epsilon_{\rm sr}
\,,\qquad
\epsilon_2\neq 4\epsilon_{\rm sr} - 2\eta_{\rm sr}
\ .
\ee
In ultra-slow-roll the potential is so flat that the acceleration dominates over the pulling force of the gradient of the potential. In other words, in ultra-slow-roll the Hubble friction term is compensated for by the field deceleration,
\be
\ddot\phi+3 H\dot\phi=-V'\simeq 0
\ ,
\ee
and therefore $\dot\phi\propto a^{-3}$, leading to $\epsilon_2\simeq -6$.

Although slow-roll and ultra-slow-roll evolutions are fundamentally different, the power spectrum of primordial scalar (linear) curvature fluctuations $\cal P$, defined in the perturbed metric
\be
ds^2=-dt^2+a(t)^2 (1+2\Psi(t, {\vec x}))d\vec{x}\cdot d\vec{x}\, ,
\label{metric:2}
\ee
where $\Psi$ is equal to the gauge invariant curvature perturbation $\zeta$ in the comoving gauge $\delta\phi=0$, is given by the same formula in both scenarios, {\it i.e.}~\cite{ultra},
\be
{\cal P} = \frac{H^2}{8\pi^2\epsilon M_p^2}
\,,
\label{Power spectrum}
\ee
while the Hubble constant evolution is still dominated by the potential energy, {\it i.e.} $H^2\propto V$.
We then see that, at least for the slow-roll case, the flatter is the potential, the larger is the power spectrum.

Let us now take this to the extreme, {\it i.e.} we aim to follow an inflationary trajectory that
starts from a dominant potential gradient, where the inflaton undergoes standard slow-roll,
to an ultra-slow-roll phase where $V'(\phi_{\rm usr})\ll \ddot\phi$.
In this case the field acceleration is small with respect to the friction during the slow-roll phase.
However, to reach the ultra-slow-roll regime the acceleration has to grow enough to overcome the potential gradient (but not too much otherwise one would exit from inflation). During this stage the velocity of the scalar, and thus the slow-roll parameter $\epsilon$, grows (is overshot from the slow-roll phase).

After that, the system transits into an ultra-slow-roll phase in which the power spectrum,
reduced by the higher value of $\epsilon$ and a lower value of the potential in the transition region,
starts exponentially increasing as $\propto e^{6 N_{\rm usr}}$, where $N_{\rm usr}=\int_{\rm usr} H dt$ is the number of e-folds
during the ultra-slow-roll phase. Given that the total number of e-foldings of inflation is bounded to $\sim 65$
(see for example~\cite{prokopec} for a discussion on this bound),  the recipe for obtaining a large power spectrum
at sub-CMB scales is to have a stage of standard inflation with as small number of e-foldings as possible and an
overshooting epoch that is as short as possible,  followed by
an ultra-slow-roll phase with the number of e-foldings as large as possible.
In what follows we show that the model discussed in~\cite{bellido} fails to satisfy these requirements.


\section{A prototype potential with an inflection point}
\label{A prototype potential with an inflection point}

Inspired by the Higgs inflation at a critical point~\cite{critical}, the authors of~\cite{bellido} have
suggested the following potential for the scalar field,%
\be\label{proto}
V(\phi)
=\frac{\lambda}{12}\phi^2 v^2 \frac{6-4a \frac{\phi}{v}+3\frac{\phi^2}{v^2}}{\left(1+b\frac{\phi^2}{v^2}\right)^2}\ ,
\ee
where for $b=1-\frac{1}{3}a^2+\frac{a^2}{3}\left(\frac{9}{2 a^2}-1\right)^{2/3}$ an inflection point is generated. In \cite{bellido} the values $a=3/2$ and $b=1$ are chosen so that the inflection point appears at $\phi_{\rm infl}=v$. In the original paper~\cite{bellido} however, $b$ was taken to slightly deviate from $1$.
The reason for doing that was the incorrect belief that at the inflection point the slow-roll parameter $\epsilon$ is
still proportional to the potential gradient and thus the power spectrum diverges there.
As discussed above however, this does not happen. Instead at the inflection point the system is ultra-slow-rolling
rather than slow-rolling and the formulae for ultra-slow-roll apply.
As we checked that the small detuning introduced in \cite{bellido} does not qualitatively change the dynamics of the system (contrary to what claimed in \cite{bellido}), we will not use this detuning in what follows\footnote{ This comment refers to the potential (\ref{proto}) with the values of parameters chosen up to the archive version 3 of \cite{bellido}. After our paper appeared on the archive, the authors of \cite{bellido}, agreeing with our analysis, have changed all values of parameters (in version 4) so that slow-roll analysis could be used. The net result is however to have a way smaller peak of the power spectrum. The reason is that, with these new values, the authors do no longer have a {\it near} inflection point. In addition, in this new version, the authors claim that they can have a peak in the power spectrum as large as they want, by opportunely de-tuning the potential. We strongly disagree with that. In fact, if the detuning is small, our analysis applies, while if it is large, slow-roll analysis applies while the peak decreases, contrary to what is claimed in \cite{bellido}.}.

\begin{figure}[t!]
\begin{center}
\includegraphics[scale=0.55]{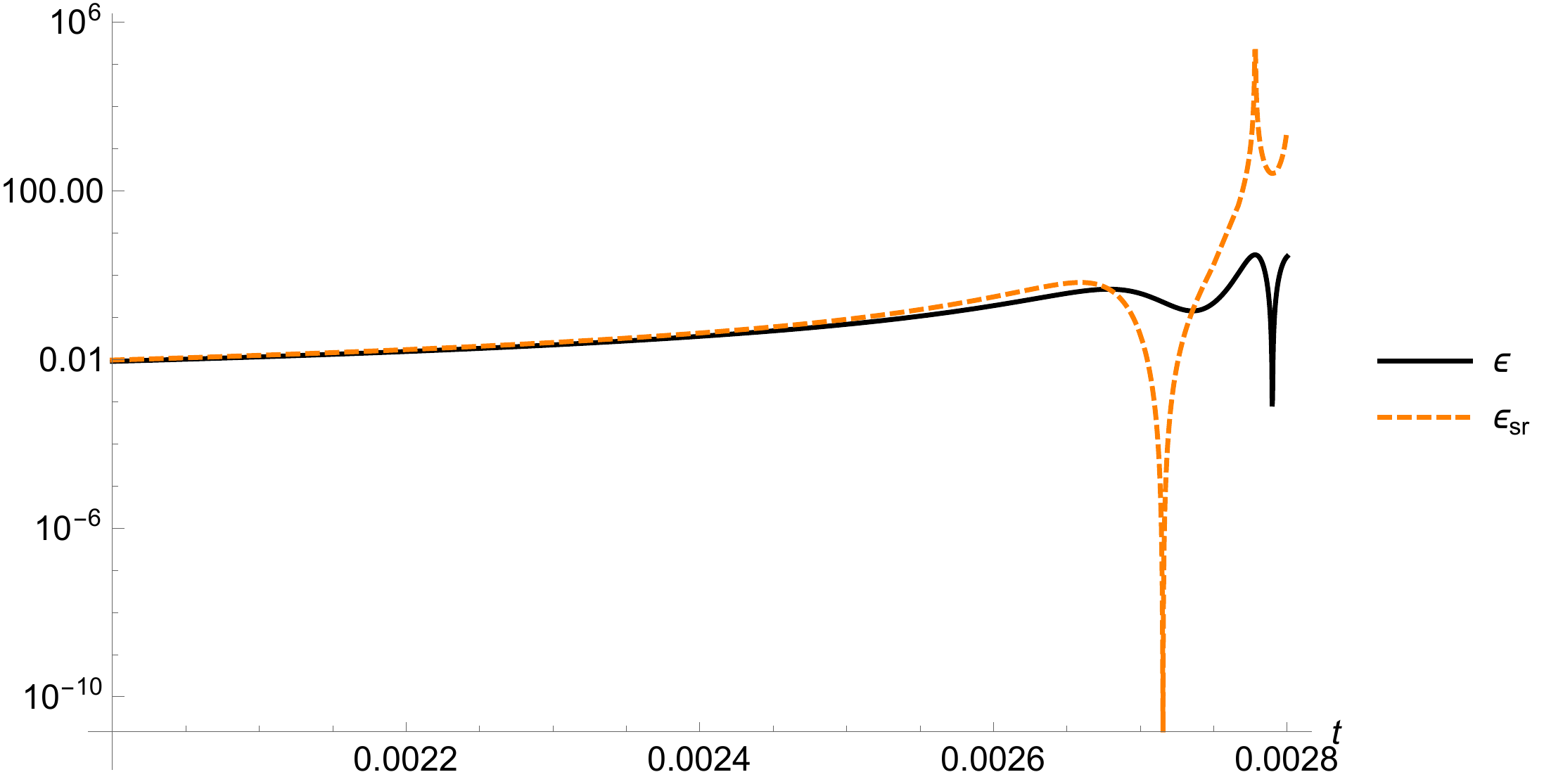}
\caption{The geometric ($\epsilon$) and approximate ($\epsilon_{\rm sr}$) principal slow-roll parameters
as a function of time (in units of $M_p^{-1}=10^{-9}$).
Notice that the slow-roll approximation breaks down before reaching the inflection point
and that $\epsilon_{\rm sr}$ vanishes at the inflection point,
$t_{\rm infl}\approx 0.00272\times10^9/M_p$,
while the geometric slow-roll parameter $\epsilon$ defined in~(\ref{epsilon:g}) remains finite at all times.
It is therefore of utmost importance to use the correct definition of $\epsilon$
 in the power spectrum~(\ref{Power spectrum}).
In this and in the subsequent figures for convenience we set $M_p=10^9$ for numerical stability.}
\label{fig1}
\end{center}
\end{figure}
\begin{figure}[t!]
\begin{center}
\includegraphics[scale=0.5]{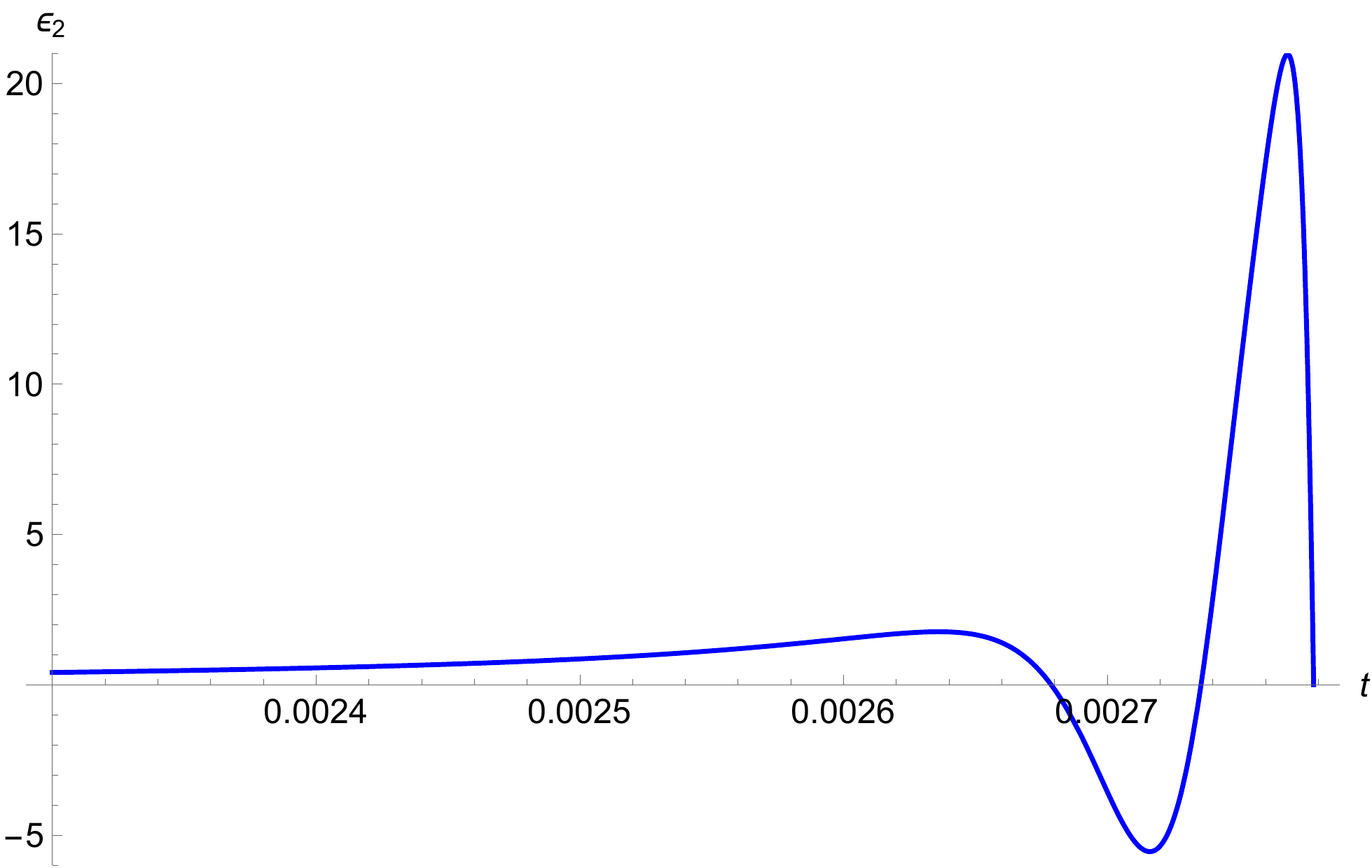}
\caption{The second geometric slow-roll parameter $\epsilon_2$ defined in~(\ref{eta:g}) as a function of time (in units of $M_p^{-1}$).
The slow-roll approximation breaks down before reaching the inflection point
($t_{\rm infl}\approx 0.00272\times 10^9/M_p$), at which $\epsilon_2$ reaches a minimum of $\epsilon_{2\,\rm min}\simeq -6$.
Inflation ends soon thereafter (the second, positive peak in $\epsilon_2$ signifies the end of inflation).}
\label{fig2}
\end{center}
\end{figure}

For a better comparison we consider here the same parameters as chosen in~\cite{bellido}.
Those are $v=\sqrt{0.121} M_p$, $\lambda=1.21\times 10^{-7}$,
and
the initial field value is, $\phi_{\rm in}\simeq 11.72\ v$~\footnote{Note that from the first to third archive version
of~\cite{bellido}, where the infection point plays a fundamental role, these parameters change. We have nevertheless checked all versions and none of them produce
the claimed spectrum.}. By choosing these  parameters one finds that,
during the slow-roll phase the geometric slow-roll parameter $\epsilon$
agrees with the slow-roll approximation $\epsilon_{\rm sr}$ excellently, as can be seen in figure~\ref{fig1}.
Near the inflection point however, the slow-roll approximation breaks down (as it is evidenced by diverging $\epsilon$ and
$\epsilon_{\rm sr}$), and the system enters an ultra-slow-roll regime,
as can be seen from the behaviour of $\epsilon_2$ in figure~\ref{fig2}.
Indeed, in figure~\ref{fig2} we see that, as expected, $\epsilon_2$ reaches a value of approximately $-6$
at the inflection point at time $t_{\rm infl}\approx 0.0027\times 10^9/M_p$.

\begin{figure}[t!]
\begin{center}
\includegraphics[scale=0.65]{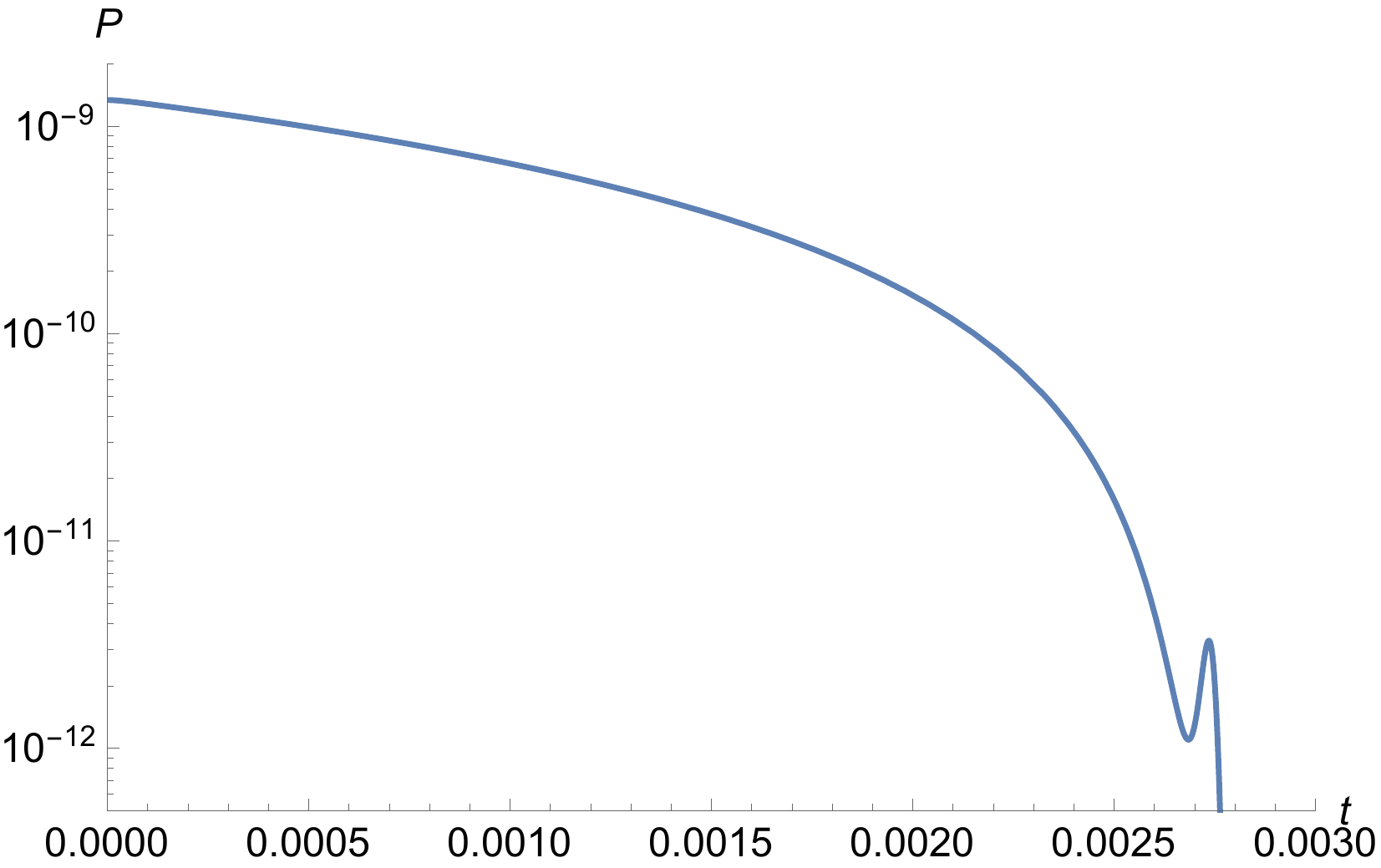}
\caption{The evolution of the power spectrum ${\cal P}$ in time with the parameters chosen as in version $3$ of~\cite{bellido}.
We see that this model exhibits a significant suppression of the power spectrum due to the overschoot, followed
 by a modest amplification due to the resonance near the inflection point.}
\label{fig3}
\end{center}
\end{figure}

The evolution of the power spectrum (in time) is shown in figure~\ref{fig3}. There we clearly see that the power spectrum gets suppressed before exponentially increasing during ultra-slow-roll. However, the number of e-foldings of the ultra-slow-roll phase is only a 
few so that the power spectrum cannot get to the quoted value of $\sim 10^{-4}$ in~\cite{bellido}.
Finally, because of the completely different dynamics analysed here and in~\cite{bellido}, the system with these parameters undergoes only about $30$ e-foldings of inflation rather than $62$ quoted in~\cite{bellido}. This is already suggesting us that there is room in the numerical choice of the parameters of the system such to increase the the number of ultra-slow-roll e-foldings. In this way, as we shall show in the next section, we will be able to recover the lost power due to the over-shoot and indeed obtain a large enhancement of the power spectrum at smaller than CMB scales.

A more interesting potential than~(\ref{proto}) is given by the Higgs potential non-minimally coupled to gravity as in the Higgs inflation of~\cite{critical}. At the near-critical point and for the parameters chosen by the authors of~\cite{bellidohiggs} we
have also checked that the same problems of the previous potential occur, {\it i.e.}
no significant amplification of the power spectrum is generated.


\section{Enhancing the power spectrum at the inflection point}

In this section, as a proof of principle, by using the potential (\ref{proto}) with appropriately tuned parameters, we
show that $\cal P$ can indeed have a huge amplification at sub-CMB scales. Fixing for comparison with \cite{bellido} a total number of e-foldings $N\simeq 62$, we will then find the maximal amplification that can be obtained from (\ref{proto}).

In order to have a large peak in the power spectrum, one has to reduce the overshooting.
This means that the region in which the standard inflation occurs should be not too far from the inflection point. In order to reduce
this distance we can maximise $b$. In addition, to provide enough e-foldings for the exponential grow during the ultra-slow-roll phase, we should also minimise $n_s$.

Firstly then, we consider the potential (\ref{proto}) with $b=1.5$, {\it i.e.}
the maximal value of $b$ such to keep an inflection point. The inflection point is then generated for $a=1/\sqrt{2}\simeq 0.707$.
Using the same set of cosmological observations as in~\cite{bellido} and considering the necessary running of the spectral index in order to obtain a peak in the power spectrum, one has, at $k_*=0.05\ {\rm Mpc}^{-1}$ and, at
$95\%$ confidence level \cite{planck},
\be
\label{limits}
\ln\left(10^{10} A_s^2\right)&=&3.094\pm 0.068\ ,\cr
n_s&=&0.9569\pm 0.0154\ ,\cr
\frac{dn_s}{d\ln k}&=&0.011\pm 0.028\ ,
\ee
where $A_s$ is the amplitude of scalar fluctuations at $k_*$.

By keeping ourselves within $2\sigma$ distance of the central values in (\ref{limits}), the peak in the power spectrum is maximised by choosing $n_s(k_*)=0.9415$, ${\cal P}(k_*)\simeq 2.06\times 10^{-9}$. \footnote{Since we start from an early slow-roll phase, we can approximate $n_s=1-6\epsilon_{sr}+2\eta_{sr}$.}
Again, for comparison, we will assume as in \cite{bellido} that the scale corresponding to $62$ e-foldings is $k_*$.

\begin{figure}[t!]
\begin{center}
\includegraphics[scale=0.5]{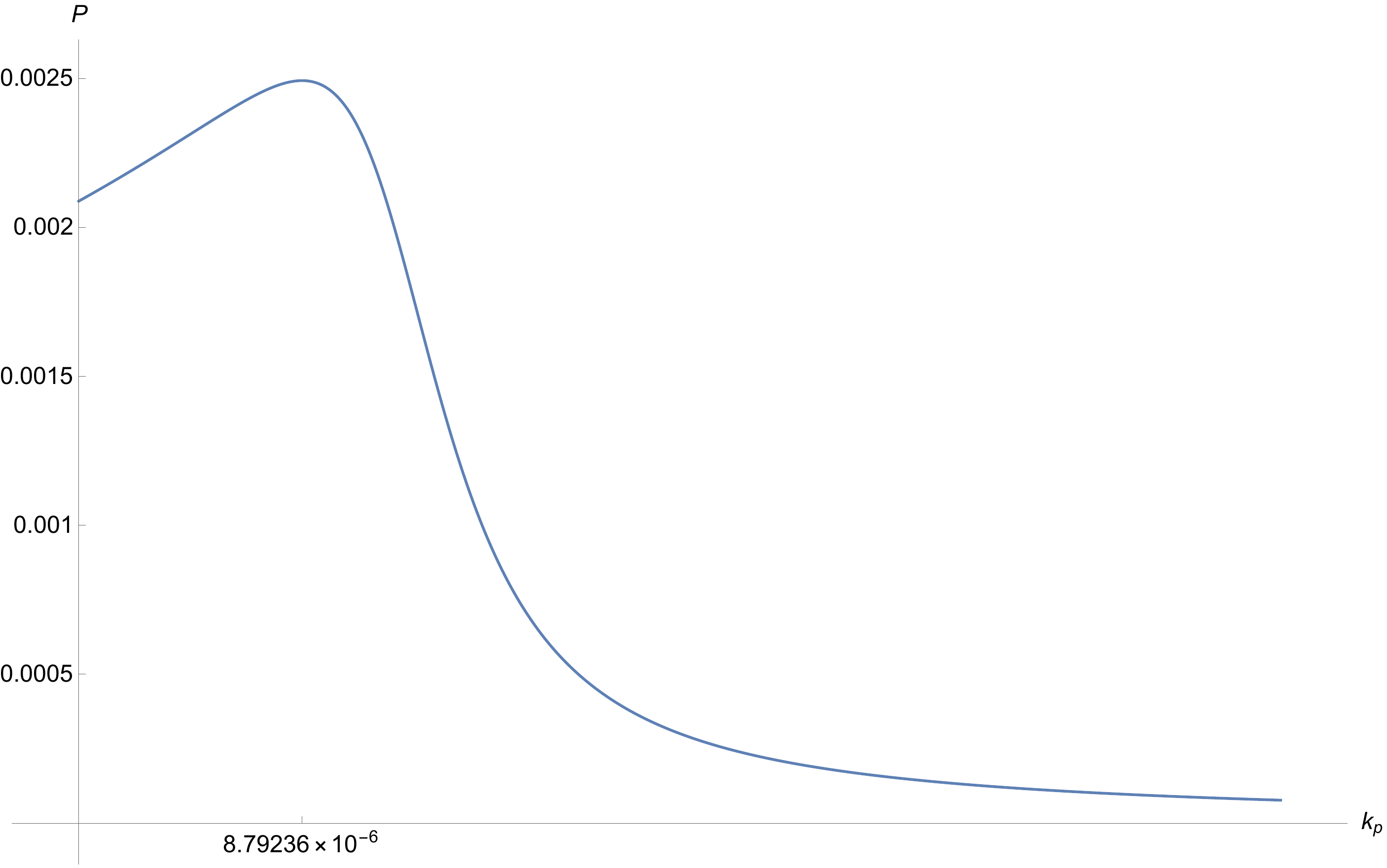}
\caption{Power spectrum as a function of the physical momentum at Hubble crossing in Planck units, {\it i.e.} $k_p/M_p\equiv\frac{k_*(t)}{a M_p}=\frac{H(t)}{M_p}$,
where $k_*(t)=aH(t)$. It is clear that the peak is very narrow and centred in $k_{p*}/M_P\simeq 8.79\times 10^{-6}$.
More precisely, in this plot: $\frac{k_p|_{\rm max}-k_p|_{\rm min}}{k_{p*}}\sim 3\times 10^{-9}$, which implies that the width of the peak is $\sim 10^{-14}M_p$.}
\label{fig4}
\end{center}
\end{figure}
\begin{figure}[t!]
\begin{center}
\includegraphics[scale=0.55]{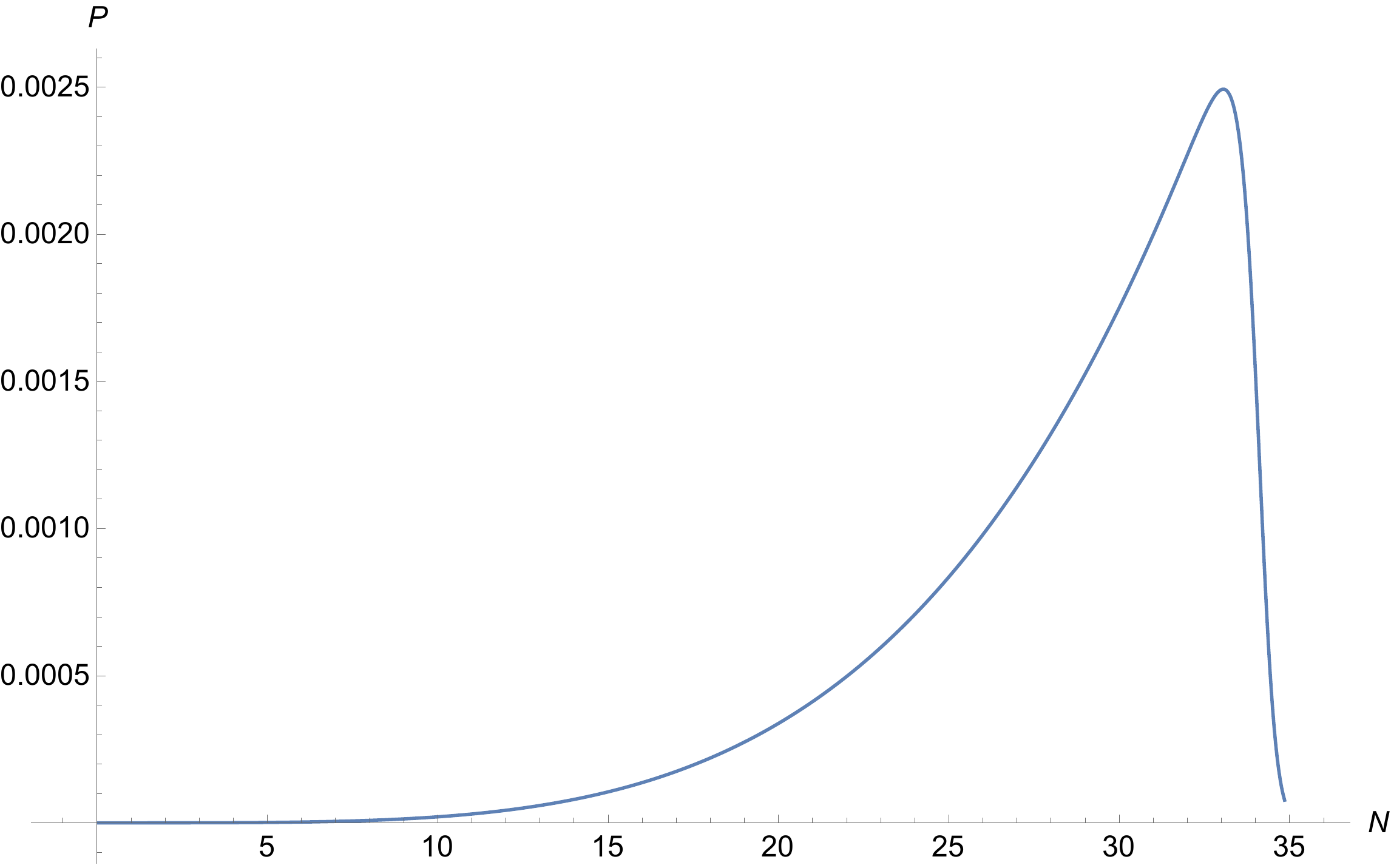}
\caption{Maximal power spectrum obtained from (\ref{proto}) with respect to the number of e-foldings from the end of inflation.}
\label{fig5}
\end{center}
\end{figure}

These values lead to $\lambda\simeq 1.86\times 10^{-6}$, $v=0.196M_p$, and for the initial scalar field value $\phi_{\rm in}\simeq 2.2 M_p$. Finally, our model gives $r\simeq0.009$ and $dn_s/d\ln k\simeq -0.002$, which amply
satisfies the current bound on the amplitude of primordial gravitational waves, $r<0.09$, and running of spectral index. With these values the power spectrum reaches a maximum of $\sim 0.0025$, which is a way larger than the maximum value of~\cite{bellido} of $\sim 7\times 10^{-5}$. A plot of the power spectrum in terms of the physical wave number ($k_p$) is given in figure~\ref{fig4}. In 
Figure~\ref{fig5} we instead plot the same potential in terms of e-foldings counted from the end of inflation. We see that, as in \cite{bellido}, the maximum is peaked around $33$ e-foldings.

\begin{figure}[t!]
\begin{center}
\includegraphics[scale=0.55]{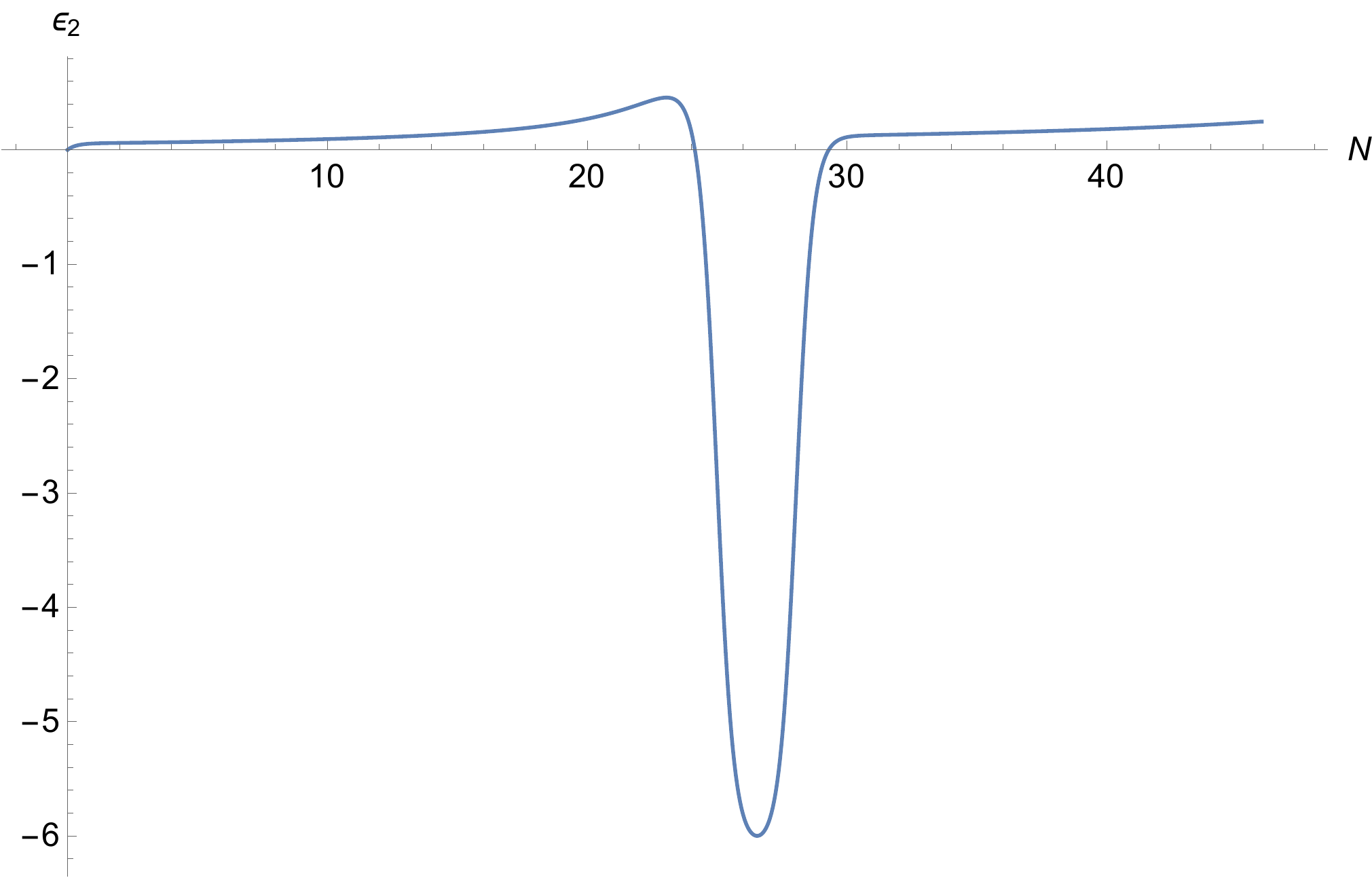}
\caption{$\epsilon_2(N)$. From this figure we clearly see that $N_{\rm usr}\sim {\rm a\ few}$}
\label{eta_N}
\end{center}
\end{figure}

Finally, the ultra-slow-roll phase is characterised by two curvature modes: a constant and a growing one \cite{ultra}. In principle the growing mode would help in the amplification of the potential, however, as the ultra-slow-roll phase only last for few e-foldings, see figure~\ref{eta_N}, the constant mode actually dominates the power spectrum.

\subsection{Is it enough for PBH?}

The density of primordial black holes with respect to the total density of the universe at the formation time is \cite{florian}
\be
\beta=\frac{\rho_{PBH}}{\rho_{tot}}\simeq\frac{1}{2}{\rm erfc}\left(\frac{\delta_c}{\sqrt{2{\cal P}_{\delta}}}\right)\ ,
\ee
where $\delta_c$ is the critical density perturbation triggering gravitational collapse from scalar modes re-entering the
horizon (Hubble volume) during radiation epoch, and ${\cal P}_{\delta}=\frac{k^3}{2\pi^2}\langle|\delta_{\rm \vec k}|^2\rangle $
is the power spectrum of the density perturbation $\delta\equiv\frac{\delta\rho}{\rho}$, where $\rho$ is the matter density.

In \cite{bellido} it has been assumed that $\delta\sim \zeta$. However, the two are related by an important numerical factor \footnote{CG would like to thank Jaume Garriga for pointing this out.}. During radiation we have \cite{misao2} $\delta=\frac{4}{9}\zeta$. The factor $4/9$ is easy to understand. In Newtonian gauge the Poisson equation provides a factor $2/3$ between the density contrast and the Newtonian potential. Another factor $2/3$ comes from treading the Newtonian potential for $\zeta$.

If, as in our case, the large peak in the curvature perturbation spectrum is very narrow, one then finds that \cite{misao2}
\be
\beta\simeq\frac{1}{2}{\rm erfc}\left(\frac{\zeta_c}{\sqrt{2{\cal P}}}\right)\ .
\ee
In addition, Ref.~\cite{bellido} used the rescaled range of $\delta_c$ found in~\cite{harada} instead of the bare ranges that are compatible with earlier works, {\it e.g.}~\cite{musco}. For this reason, assuming that PBHs are formed during
radiation domination, one should instead consider $\delta_c\simeq 0.45$~\cite{florian}, \cite{harada}, \cite{musco} and thus $\zeta_c\simeq 1.01$. 
Then, at the peak we get $\beta\sim 10^{-91}$~,~\footnote{Had we used $\delta\sim\zeta$ and $\delta_c=0.0875$, as in \cite{bellido},  we would have obtained $\beta\sim 0.04$.} which is obviously way too small to produce any non-negligible late time PBH abundances (we also checked that enlarging the number of e-foldings 
to $\sim 65$ does not help much).

This result should not be discouraging. Although the potential (\ref{proto}) does not serve for the purpose of generating a significant amount of PBH, we expect that one can always find a suitable potential with an inflection point (or better a flat direction) producing the right amplitude of scalar perturbations to match the observed abundance of Dark Matter. However, as $\beta$ is exponentially sensitive to small variations of $\cal P$, we believe that any choice of that potential should be strongly physically supported or, at the very least, very stable under quantum corrections. As our paper aims only to elucidate the correct mechanism of power spectrum enhancement via an inflection point, the prototypical potential (\ref{proto}) served our case. Thus, the quest of finding a stable model of inflation, able to generate the right density of PBH
matching current observation, is left for future research.

\section{Conclusions}

In this work we have revisited the important question of the production of primordial black holes (PBHs) by amplified
spectrum of adiabatic cosmological perturbations. We have reanalysed the fascinating claim
of the recent reference~\cite{bellido}, in which inflation near an inflection point of the inflaton potential was used to get a
resonant enhancement of the adiabatic power spectrum needed to produce enough PBHs.
We came to the conclusion that understanding the amplification near the inflection point
requires a more careful analysis than it was done in~\cite{bellido}. The main reason is that near an inflection point, slow-roll analysis breaks down, as already noticed in \cite{tallin} \footnote{See also \cite{rami} where the authors notice some discrepancies between the exact and slow-roll anaysis for certain classes of potentials with small curvatures.}. Contrary however to \cite{tallin}, where slow-roll was then enforced by appropriately deforming a potential with an inflection point, here we have studied the case in which the inflection point is kept so to generate a large peak in the power spectrum.

Our main result can be summarised as follows:
near the inflection point the inflaton enters an ultra-slow-roll regime, during which
the standard slow-roll methods (used in~\cite{bellido}) break down. However, the methods of ultra-slow-roll
give a clear picture of why there is a resonant enhancement. Namely, near the inflection point $V^\prime(\phi)$
almost vanishes and the inflaton enters into a strongly decelerating phase during which $\dot \phi\propto 1/a^3$.
This then results in a strong reduction of the principal (geometric) slow-roll parameter, $\epsilon_2\propto a ^{-6}$,
and consequently to a large amplification of the adiabatic power spectrum by a factor,
 $P\propto e^{6N_{\rm usr}}$, where $N_{\rm usr}$ is the number of e-foldings that the system spends in ultra-slow-roll.

 We have shown that, with
a suitable choice of the parameters, one can get a large amplification of the power spectrum
(see figure~\ref{fig4}). However, that is unfortunately not enough
to produce the required amount of PBHs needed to explain dark matter.

One would very much like to understand under what general conditions one gets a strong amplification of the power spectrum.
We have performed an in-depth analysis of that question and came to the following conclusion.
The resonant amplification will generally be stronger if:
\begin{enumerate}
\item[$\bullet$] {\it One reduces the amount of overshooting.} This can be achieved e.g. by pushing
the resonance to earlier times in inflation, {\it i.e.} closer to the smallest observed CMB scales.
\item[$\bullet$] {\it One increases the period of ultra-slow-roll}. This can be generally achieved by flattening of
the potential. A practical way of doing that is to put as close as possible the slow-roll region (in the potential) to the inflection point and to minimise the scalar spectral index $n_s$.
\end{enumerate}

The enhancement of the power spectrum shown in figure~\ref{fig4} is obtained by tuning of the parameters. We believe this is bound to be a general feature of any enhancements of the power spectrum via an inflection point.

An important question is how large the required fine tuning actually is. Any attempt to answer that question here would
necessarily address that question for a specific potential. While the amount of fine tuning
required to get DM from PBHs may be large, we do not know how large
fine tuning is generically, {\it i.e.} without specifying the potential.
This may be ever the wrong question to ask as there may be
a physical mechanism that naturally generates potentials that possess the desired strong resonance.
A more detailed investigation of this fascinating question, and a suitable potential able to generate an amplitude of scalar perturbations high enough in order to interpret primordial black holes as dark matter candidates,  is left for future work.

An interesting paper~\cite{Pattison:2017mbe} appeared after ours in which the authors argue that stochastic effects on the inflaton dynamics and non-Gaussianities, might be important. In particular -- in accordance with~\cite{Young:2015cyn} -- the authors of~\cite{Pattison:2017mbe} point out that, when this is the case, the non-Gaussian tail generated by stochastic effects can contribute most of the PBHs production. These results however need further confirmation in any region violating slow-roll, as it is on and around an inflection point (ultra-slow-roll case). We leave this study for future research.

\section*{Acknowledgements}
CG would like to thank Jaume Garriga, Jordi Miralda-Escud\'e, Ilia Musco, Florian Kh\"unel and Alvise Raccanelli, for illuminating discussions.
CG is supported by the Ramon y Cajal program and
partially supported by the Unidad de Excelencia Maria de Maeztu Grant No. MDM-2014-0369 and FPA2013-
46570-C2-2-P grant.

\section*{References}

\bibliography{bibliography}

\end{document}